\documentclass[aps,prd,twocolumn,showpacs,superscriptaddress,groupedaddress,nofootinbib]{revtex4}  % for review and submission
\usepackage{graphicx}  % needed for figures
\usepackage{dcolumn}   % needed for some tables
\usepackage{bm}        % for math
\usepackage{amsmath,amssymb}   % for math
\usepackage{aas_macros}
\usepackage{multirow}
\usepackage{color}
\usepackage{verbatim}
\usepackage{url}
\usepackage{hyperref}

\newcommand{\mr}{\mathrm}

\newcommand{\bea}{\begin{eqnarray}}
\newcommand{\eea}{\end{eqnarray}}

\begin{document}
% The following information is for internal review, please remove them for submission
\widetext
% the following line is for submission, including submission to the arXiv!!
%\hspace{5.2in} \mbox{Fermilab-Pub-04/xxx-E}

\title{Primordial density and BAO reconstruction}

\author{Hong-Ming Zhu}
\affiliation{Key Laboratory for Computational Astrophysics, National Astronomical Observatories, Chinese Academy of Sciences, 20A Datun Road, Beijing 100012, China}
\affiliation{University of Chinese Academy of Sciences, Beijing 100049, China}

\author{Ue-Li Pen}
\affiliation{Canadian Institute for Theoretical Astrophysics, University of Toronto, 60 St. George Street, Toronto, Ontario M5S 3H8, Canada}
\affiliation{Dunlap Institute for Astronomy and Astrophysics, University of Toronto, 50 St. George Street, Toronto, Ontario M5S 3H4, Canada}
\affiliation{Canadian Institute for Advanced Research, CIFAR Program in Gravitation and Cosmology, Toronto, Ontario M5G 1Z8, Canada}
\affiliation{Perimeter Institute for Theoretical Physics, 31 Caroline Street North, Waterloo, Ontario, N2L 2Y5, Canada}

%\author{Matthew McQuinn}
%\affiliation{Department of Astronomy, University of Washington, Seattle, WA 98195, USA}

\author{Xuelei Chen}
\affiliation{Key Laboratory for Computational Astrophysics, National Astronomical Observatories, Chinese Academy of Sciences, 20A Datun Road, Beijing 100012, China}
\affiliation{University of Chinese Academy of Sciences, Beijing 100049, China}
\affiliation{Center of High Energy Physics, Peking University, Beijing 100871, China}

\date{\today}

\begin{abstract}
We present a new method to reconstruct the primordial (linear) density field
using the estimated nonlinear displacement field.
The divergence of the displacement field gives the reconstructed density field.
We solve the nonlinear displacement field in the 1D cosmology and show the 
reconstruction results. 
The new reconstruction algorithm recovers a lot of linear modes 
and reduces the nonlinear damping scale significantly.
The successful 1D reconstruction results imply the new algorithm should also 
be a promising technique in the 3D case.
\end{abstract}

\pacs{}
\maketitle

%\section{\label{sec:level1}First-level heading}
% sections are not used for PRL papers

\section{Introduction}
The observed large-scale structure of the Universe, which contains a wealth of 
information such as the nature of dark energy, neutrino masses, and primordial
power spectrum etc, is a powerful probe of cosmology.
The matter power spectrum has been measured to significant accuracy in the 
current galaxy surveys and the precision will continue to improve with future 
surveys. 
However, the nonlinear gravitational evolution is a complicated process and 
makes it difficult to model the small-scale inhomogenities. This has led to
many theoretical challenges in developing perturbation theories 
(see e.g. \cite{2016matt} for a brief review).
On the other hand, various reconstruction techniques have been proposed to 
reduce nonlinearities in the density field, in order to obtain better statistics
\cite{2007bao,2015PhRvD..92l3522S}. 
 
The standard BAO reconstruction uses the negative
Zel'dovich (linear) displacement to reverse the large-scale bulk flows 
\cite{2007bao}. 
The nonlinear density field is usually smoothed on the linear scale 
($\sim10\ \mr{Mpc}/h$) to make the Zel'dovich approximation valid.
Actually, the fully nonlinear displacement which describes the motion beyond
the linear order (the Zel'dovich approximation) can be solved from the nonlinear
density field. 
While the algorithm is complicated in the three spatial dimensions, it is quite simple in the 1D case, which is basically the ordering of mass elements 
(sheets).
The 1D cosmological dynamics corresponds to the interaction of infinite sheets
of matter where the force is independent of distance \cite{2016matt}.
%The sheets are moving in a Hubble flow relative to one another and the surface
%density of each sheet scales as $a^{-2}$.
The simplified 1D dynamics provides an excellent means of understanding the 
structure formation and testing perturbation theories \cite{2016matt}.
In this paper we solve the fully nonlinear displacement in 1D and present
a new method to reconstruct the primordial density field and hence the linear 
BAO information.

This paper is organized as follows. 
In Section \ref{rec}, we present the reconstruction algorithm in the 1D case. 
In Section \ref{sim}, we briefly describe the 1D $N$-body simulation. 
In Section \ref{res}, we show the results of reconstruction.
In Section \ref{dis}, we discuss the 3D case and future improvements.
%=======================================

\section{Reconstruction algorithm}
\label{rec}
The Lagrangian displacement 
${\Psi}({q})$ fully describes the motion of mass elements.
The Eulerian position ${x}$ of a mass element is
\bea
{x}={q}+{\Psi}({q}),
\eea
where ${q}$ is the initial Lagrangian position of this mass element.
In simulations, mass elements (sheets) are labeled by their initial 
Lagrangian coordinates. Once we know their Eulerian positions, the displacement 
field is obtained. However, in observations we only have the unlabelled 
Eulerian coordinates. 
The estimated displacement at the Lagrangian coordinate $q=iL/N$ is 
\bea
s(q)=x_i-iL/N,
\eea
where we have ordered the sheet lables $i$ from left to right, $L$ is the box
size, and $N$ is the sheet number. 
Here, $q=iL/N$ is the estimated initial Lagrangian position for the $i$th sheet
at position $x_i$.
If no shell crossing happens, the reconstructed displacement is exact up to
a global shift. In the nonlinear regime once shell crossing occurs, the 
estimated displacement field is quite noisy on the scale $\sim L/N$.
To reduce stochasticities in the estimated displacement field, we can use the
averaged displacement of $n_p$ particles 
\bea
s(q)=\frac{1}{n_p}\sum_{j=i}^{i+n_p-1}x_{j}-in_pL/N,
\eea
where $q=in_pL/N$ and $j$ is the sheet label. 
Here $i$ varies from $0$ to $N/n_p$ and $j$ varies from $0$ to $N$.
We take $n_p=5$ to estimate the displacement field in this paper.

The derivative (actually the divergence) of $s(q)$ gives the reconstructed 
density field 
\bea
\delta_r({q})=-\frac{\partial s(q)}{\partial q},
\eea
i.e., the differential motion of mass elements. 
Reconstruction from the gridded density field can be implemented following
the same principle, which we adopt in the following calculations.
%=======================================

\section{Simulations}
\label{sim}
The 1D $N$-body dynamics can be simulated using the particle-mesh (PM) method.
The 1D simulations we use are run with the 1D PM code in Ref. \cite{2016matt}.
The 1D simualtion involves $3\times10^8$ sheets with $3\times10^8$ PM elements 
in a $10^8\ \mr{Mpc}$ box. The 1D simulation assumes a matter-dominated 
background cosmology ($\Omega_m=1$) and have the same dimensionless power 
spectrum as the concordance cosmology, i.e., 
\bea
kP_\mr{1D}(k)/\pi=k^3P_\mr{3D}(k)/(2\pi^2),
\eea
where $P_\mr{3D}$ is the 3D linear power spectrum from the linear Boltzman code.

The initial condition is generated using the Zel'dovich approximation.
Since the Zel'dovich approximation is exact up to shell crossing, the PM 
calculation is started at $z=10$. In the analysis, we use the output at $z=0$.
We also scale the initial density field by the linear growth factor to get the
linear density field at $z=0$. 

Note that the nonlinear evolution in 1D is more significant than the 3D case.
The nonlinear evolution in the concordance (3D) cosmology at $z=0$ is 
only comparable to the 1D cosmology at $z=1$ \cite{2016matt}.
%=======================================

\section{Results}
\label{res}
\begin{figure}[tbp]
\begin{center}
\includegraphics[width=0.48\textwidth]{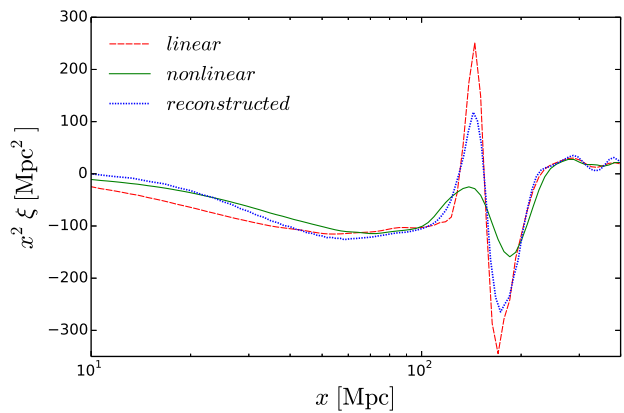}
\end{center}
\vspace{-0.7cm}
\caption{The linear (dashed line), nonlinear (solid line), and reconstructed 
(dotted line) correlation functions.  
The distortion of the BAO peak is reduced by reconstruction.}
\label{fig:ps}
\end{figure}

Figure \ref{fig:ps} shows the linear, nonlinear and reconstructed correlation
functions.
Since the BAO feature in 1D is sharper than that in 3D, the smearing of the BAO
peak in 1D is more subtantial \cite{2016matt}. 
Nevertheless, the new reconstruction method sharpens the peak signigicantly.
The nonlinear density field $\delta(x)$ is given on the Eulerian position $x$,
while the reconstructed density field $\delta_r(q)$ is calculated on the 
Lagrangian position $q$. 

\begin{figure}[tbp]
\begin{center}
\includegraphics[width=0.48\textwidth]{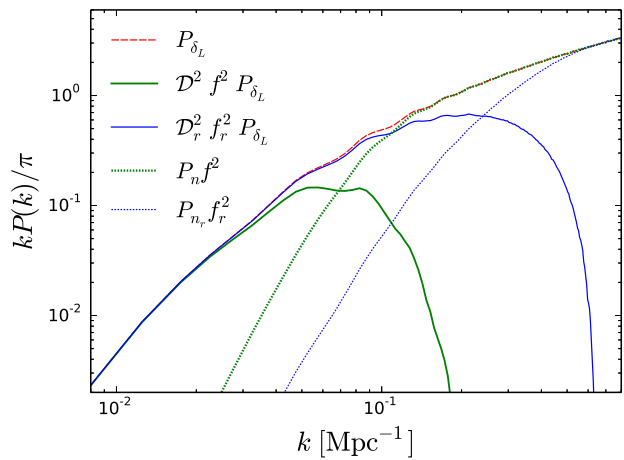}
\end{center}
\vspace{-0.7cm}
\caption{The linear power spectrum (dashed line), the linear parts of the
nonlinear (thick solid line) and reconstructed (thin solid line) power spectra,
the noise parts of the nonlinear (thick dotted line) and reconstructed
(thin dotted line) power spectra. 
For visual comparisions, we rescale both the linear and noise parts by 
$f^2=P_{\delta_L}/P_{\delta}$ and $f^2_r=P_{\delta_L}/P_{\delta_r}$ for the nonlinear and reconstructed fields, respectively. 
The noise terms dominate over the signals at 
$k\gtrsim0.07\ \mr{Mpc}^{-1}$ for the nonlinear field and $k_q\gtrsim0.24\ \mr{Mpc}^{-1}$
for the reconstructed field.}
\label{fig:pn}
\end{figure}

To conveniently quantify the linear information $\delta_L$ in 
the nonlinear density field $\delta$, we decompose the nonlinear density field
$\delta$ as
\begin{eqnarray}
\delta({k})=b({k})\delta_L({k})+n({k}).
\end{eqnarray}
Here, $b\delta_L$ is completely correlated with the linear density field 
$\delta_L$. Correlating the nonlinear density field with the linear density 
field,  
\bea
\langle\delta(k)\delta_L(k)\rangle=b(k)\langle\delta_L(k)\delta_L(k)\rangle,
\eea
we obtain 
\bea
b(k)=\frac{P_{\delta\delta_L}(k)}{P_{\delta_L}(k)}.
\eea
Nonlinear evolution drives $b(k)$ to drop from unity, reducing the linear 
signal. Separating the part correlated with the linear density field, we have
$n(k)=\delta(k)-b(k)\delta_L(k)$.
$n(k)$ is generated in the nonlinear evolution and thus uncorrelated with
the linear density field $\delta_L$, further reducing $b\delta_L$ with respect
to $\delta$. This part induces noise in the measurement of BAO. 
Such decomposition helps to write the nonlinear power spectrum as
\bea
P_\delta(k)=\mathcal{D}(k)P_{\delta_L}(k)+P_{n}(k),
\eea
where $\mathcal{D}(k)=b^2(k)$ is the nonlinear damping factor.
Here, $b(k)$ is often referred as the ``propagator'' and $P_{n}$ is usually
called the ``mode-coupling'' term \cite{2006crocce,2008crocce,2008matsubara}.
For the reconstructed field $\delta_r(q)$, we also have
\bea
\delta_r(k_q)=b_r(k_q)\delta_L(k_q)+n_r(k_q),
\eea
where $b_r(k_q)={P_{\delta_r\delta_L}(k_q)}/{P_{\delta_L}(k_q)}$.
Similarly, the reconstructed power spectrum is given by
\bea
P_{\delta_r}(k)=\mathcal{D}_r(k)P_{\delta_L}(k)+P_{n_r}(k),
\eea
where $\mathcal{D}_r(k)=b^2_r(k)$.
Here, the subscript ``$q$'' denotes that the reconstructed field is given on the
Lagrangian coordinate.
In Fig. \ref{fig:pn}, we plot the linear components and the noise terms of
the nonlinear and reconstructed fields.

\begin{figure}[tbp]
\begin{center}
\includegraphics[width=0.48\textwidth]{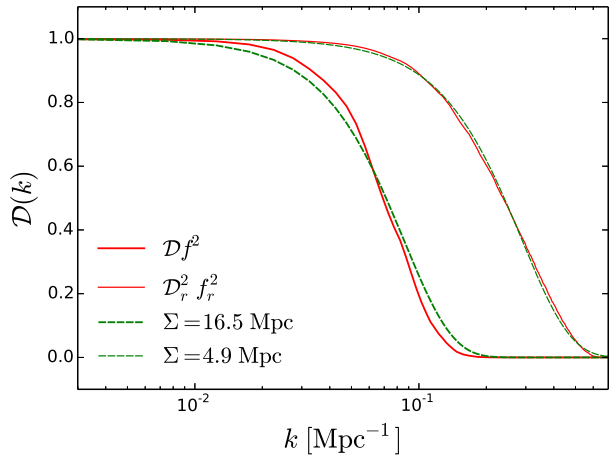}
\end{center}
\vspace{-0.7cm}
\caption{The damping factors for the nonlinear (thick solid line) and 
reconstructed (thin solid line) fields. The Gaussian BAO damping models with 
$\Sigma=16.5\ \mr{Mpc}$ (thick dashed line) and $\Sigma=4.9\ \mr{Mpc}$ 
(thin dashed line).}
\label{fig:damp}
\end{figure}

Figure \ref{fig:damp} shows the damping factors for the 
nonlinear and reconstructed fields. The damping of the linear power spectrum is 
significantly reduced after reconstruction. We also overplot the best-fitting 
Gaussian BAO damping model,
\bea
\mathcal{D}(k)=\mr{e}^{-k^2\Sigma^2/2},
\eea
with $\Sigma=16.5\ \mr{Mpc}$ and $4.9\ \mr{Mpc}$ for the nonlinear and 
reconstructed fields. 
The new BAO reconstruction algorithm reduces the nonlinear damping
scale $\Sigma$ by 70 percent. 
The damping factor for the reconstructed field is above $0.9$ for 
$k\lesssim0.1\ \mr{Mpc}^{-1}$. 
However, the 100 percent reconstruction, cancelling any nonlinear effects,
is still unachievable, as some information has been irreversibly lost. 

\begin{figure}[tbp]
\begin{center}
\includegraphics[width=0.48\textwidth]{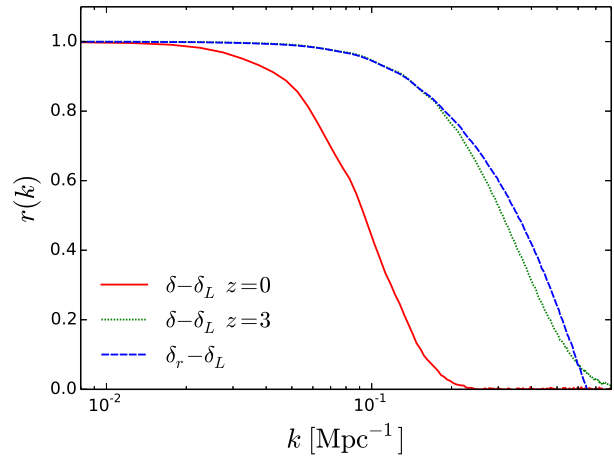}
\end{center}
\vspace{-0.7cm}
\caption{The $\delta-\delta_L$ correlation coefficients at $z=0$ (solid
line) and $z=3$ (dotted line), as well as the $\delta_r-\delta_L$ 
correlation coefficient (dashed line).}
\label{fig:xcc}
\end{figure}

Reconstruction reduces the nonlinear damping $\mathcal{D}(k)$ as well as the 
noise term $P_{n}(k)$. To quantify the overall performance, we can use the 
cross-correlation coefficient 
\bea
r(k)=\frac{P_{\delta\delta_L}(k)}
{\sqrt{P_{\delta}(k)P_{\delta_L}(k)}}
=\frac{1}{\sqrt{1+\eta(k)}},
\eea
where $\eta=P_n/(b^2P_{\delta_L})$ quantifies the relative amplitude
of $n$ with respect to $b\delta_L$. We plot the cross-correlation coefficients
in Fig. \ref{fig:xcc}. The correlation of $\delta_r$ with
$\delta_L$ is even better than that of $\delta$ at $z=3$.

\begin{figure}[tbp]
\begin{center}
\includegraphics[width=0.48\textwidth]{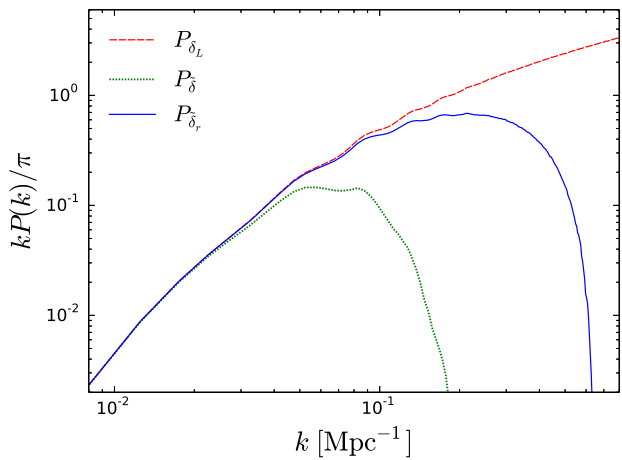}
\end{center}
\vspace{-0.7cm}
\caption{The power spectra for the linear (dashed line), filtered nonlinear
(dotted line) and filtered reconstructed (solid line) fields.
The wiggles in the reconstructed power 
spectrum are much more apparent than the nonlinear power spectrum.}
\label{fig:wf}
\end{figure}

The raw reconstructed field $\delta_r$ is still noisy on small scales 
($k_q\gtrsim0.24\ \mr{Mpc}^{-1}$). To optimally filter out the noise from the 
raw reconstructed field, we use the Wiener filter
\bea
W_r(k_q)=\frac{P_{\delta_L}(k_q)}{P_{\delta_L}(k_q)+P_{n_r}(k_q)/b_r^2(k_q)}.
\eea
Deconvolving $b_r$ and using the Wiener filter, we obtain the optimal 
reconstructed field,
\bea
\tilde{\delta}_r(k_q)=\frac{\delta_r(k_q)}{b_r(k_q)}W_r(k_q).
\eea
The power spectrum of the optimal reconstructed field $\tilde{\delta}_r$ is 
given by
\bea
P_{\tilde{\delta}_r}(k_q)=W_r^2(k_q)P_{\delta_L}(k_q)+W_r^2(k_q)
{P_{n_r}(k_q)}/{b_r^2(k_q)}.
\eea
The raw nonlinear field $\delta$ is also filtered. 
In Fig. \ref{fig:wf}, we plot the power spectra of the optimal filtered 
reconstructed and nonlinear fields. The wiggles in the reconstructed power 
spectrum are much more apparent than the nonlinear power spectrum.

\begin{figure}[tbp]
\begin{center}
\includegraphics[width=0.48\textwidth]{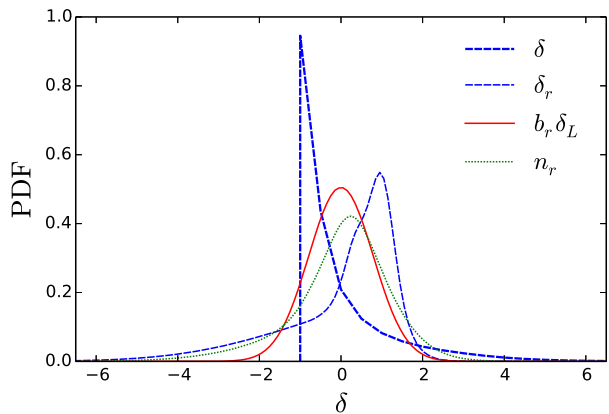}
\end{center}
\vspace{-0.7cm}
\caption{The probability distribution functions of the nonlinear (thick dashed 
line) and reconstructed (thin dashed line) fields. 
We also show the PDFs of the linear component 
(solid line) and the noise part (dotted line). The PDFs are evaluated on 
$N_\mr{grid}=6\times10^7$ grids, i.e., the grid scale is $5/3\ \mr{Mpc}$.}
\label{fig:pdf}
\end{figure}

The density fluctuation probability distribution function (PDF) quantifies the 
Gaussianity of the density field. Figure \ref{fig:pdf} shows the PDFs of
the nonlinear and reconstructed density fields. We also plot the PDFs of the
linear component $b_r\delta_L$ and the noise part $n_r$ of the reconstructed 
density field $\delta_r$. Of course the linear component is Gaussian, while 
the noise part is nonGaussian. As a result, the reconstructed density field is 
also nonGaussian. The raw nonlinear density field is clearly nonGaussian.
%=======================================

\section{ Discussions}
\label{dis}
The new reconstruction method successfully recovers the lost linear information 
on the mildly nonlinear scales (till $k\lesssim0.24\ \mr{Mpc}^{-1}$).
The result in 1D provides an intuitive view of the algorithm and motivates us to
develop the reconstruction method in 3D. The nonlinear displacement beyond the
Zel'dovich approximation in 3D can be solved using the multigrid
iteration scheme \cite{1995ApJS..100..269P}. The algorithm 
for solving the 3D nonlinear displacement is originally introduced for the 
adaptive particle-mesh $N$-body code \cite{1995ApJS..100..269P} 
and the moving mesh hydrodynamic code \cite{1998ApJS..115...19P}.
The 3D case is also more complicated since the 3D displacement field involves
a curl part (vorticity) which is generated after shell crossing, 
while this does not happen after particles cross over in 1D. 
This requires us to quantify the effect of vorticity, which can be accomplished
using $N$-body simulations. 
By decomposing the simulated displacement field into a irrotational part and
a curl part, we can study the statistical properties of different components
\cite{2013PhRvD..87f3526Z,2013PhRvD..88j3510Z}.
These will be presented in future.

The reconstructed nonlinear displacement field is also important for the 
current BAO reconstruction \cite{2007bao}, where the linear continuity
equation is adopted to solve the displacement under the Zel'dovich 
approximation. 
However, the nonlinear displacement retains much more information, describing 
the motion of dark matter fluid elements beyond the linear order.
The reconstructed displacement field $s(q)$ is given on the Lagrangian 
coordinate instead of the final Eulerian coordinate. 
This helps to correct the effect due to the use of $s(x)$ instead of $s(q)$ 
in the BAO reconstruction \cite{2015MNRAS.450.3822W,2015PhRvD..92l3522S}.
As more nonlinear effects will be removed using the nonlinear displacement,
we expect the modeling of the reconstructed density field will be simplified.

The Wiener filter is optimal for the case both the signal and the noise are
Gaussian random fields. In Fig. \ref{fig:pdf}, the PDFs of the reconstructed
density field and the noise are apparently nonGaussian.
The reconstruction can be further improved using the nonlinear filter
rather than the Wiener filter \cite{1999RSPTA.357.2561P}.
We plan to study this in future.
%=======================================

\section{Acknowledgements}
We are very grateful to Matthew McQuinn for providing the 1D $N$-body 
simulations and helpful comments on the manuscript.
We also thank Yu Yu, Tian-Xiang Mao and Wen-Xiao Xu for useful discussions.
We acknowledge the support of the Chinese MoST 863 program under Grant 
No. 2012AA121701, the CAS Science Strategic Priority Research Program 
XDB09000000, the NSFC under Grant No. 11373030, IAS at Tsinghua University, 
 and NSERC.
The Dunlap Institute is funded through an endowment established by the David Dunlap family and the University of Toronto.
Research at the Perimeter Institute is supported by the Government of Canada
through Industry Canada and by the Province of Ontario through the Ministry of
Research $\&$ Innovation.

\bibliographystyle{apsrev}
\bibliography{1d}

\end{document}